\documentclass[aps,showpacs,twocolumn,superscriptaddress]{revtex4}

\usepackage{graphicx}  
\usepackage{dcolumn}
\usepackage{amsmath}

\newcommand{\sfrac}[2]{\mbox{\footnotesize $\displaystyle \frac{#1}{#2}$}}

\begin{document}
% You should use BibTeX and apsrev.bst for references
%\bibliographystyle{apsrev}

\preprint{ANL-PHY-9800-TH-2001, UNITU-THEP-04/2001}

\title{\hspace*{\fill}{\small To appear in Phys. Rev. C}\\
Neutron electric dipole moment: \\Constituent-dressing and
compositeness}
% Optional argument for running titles on pages
%\title[]{}


\author{M.B.~Hecht}
%\email[]{hecht@theory.phy.anl.gov}
%\homepage[]{http://www.phy.anl.gov/theory/staff/mh.html}
\author{C.D.~Roberts}
%\email[]{cdroberts@anl.gov}
%\homepage[]{http://www.phy.anl.gov/theory/staff/cdr.html}
\affiliation{Physics Division, Bldg 203, Argonne National
Laboratory, Argonne Illinois 60439-4843}
%\thanks{}
\author{S.M.~Schmidt}
%\email[]{basti@pion20.tphys.physik.uni-tuebingen.de}
%\homepage[]{http://www.phy.anl.gov/theory/staff/basti.html}
\affiliation{Institut f\"ur Theoretische Physik, Universit\"at T\"ubingen, Auf
der Morgenstelle 14, D-72076 T\"ubingen, Germany}


%\date{To appear in Phys. Rev. C}

\begin{abstract}
\rule{0ex}{3.0ex} 
Contributions to the neutron's EDM, $d_n$, are calculated using a
well-constrained {\it Ansatz} for the nucleon's Poincar\'e covariant Fadde'ev
amplitude.  The momentum-dependent quark dressing amplifies the contribution
from the current-quarks' EDMs; and dressed-quark confinement and binding make
distinguishable the effect of the two CP and T violating interactions: $i
\gamma_5\sigma_{\mu\nu}\,(p_1-p_2)_\nu$ and $\gamma_5 (p_1+p_2)_\mu$, where
$p_{1,2}$ are the current-quarks' momenta.  The value of $|d_n|$ obtained
using the current-quark EDMs generated by a minimal three Higgs doublet model
of spontaneous CP violation is close to the current experimental upper
bound.
\end{abstract}
%\keywords{}

\maketitle

\section{Introduction}
\label{introduction}
The action for any local quantum field theory is invariant under the
transformation generated by the antiunitary operator CPT, which is the
product of the inversions: C - charge conjugation; P - parity transformation;
and T - time reversal.  The combined CPT transformation provides a rigorous
correspondence between particles and antiparticles, and it relates the ${\cal
S}$-matrix for any given process to its inverse process, where all the spins
are flipped and the particles replaced by their antiparticles.  Lorentz and
CPT symmetry together have many consequences, among them, that the mass and
total width of any particle are identical to those of its antiparticle.

The decay of the CP-odd eigenstate $K^0_L$ into a CP-even $2 \pi$ final state
demonstrates that the product of only C and P is not a good symmetry of the
standard model.  This entails that time reversal invariance must also be
violated and that too has been observed in detailed studies of the neutral
kaon system~\cite{Tviol}.  The separate violation of CP and T invariance can
be accommodated in the six-quark-flavour standard model: only five of the six
possible phases in the $3\times 3$ Cabibbo-Kobayashi-Maskawa (CKM) matrix can
be eliminated via phase rotations of the quark flavours -- rotating all the
quarks through the same phase cannot affect the CKM matrix -- leaving one CP
violating phase.  (In the four flavour theory, by contrast, the $2\times 2$
CKM matrix has three possible phases, all of which can be eliminated.)
However, some aspects of the standard model are not completely satisfactory,
notably the understanding of the magnitude of the direct CP violating
parameter $\epsilon^\prime$~\cite{epe,mishaKpipi,bertolini}.

Identifying phenomena that are inexplicable in the standard model is an
important focus of contemporary nuclear and particle physics, and while the
neutral kaon system is the archetype for CP and T violation, the standard CKM
model predicts much larger effects for $B$-mesons~\cite{wolfenstein}.  A
primary goal of the $B$-factories now being developed is to check these
predictions.  This is a new domain for testing the standard model.

However, it has long been known that the possession of an electric dipole
moment (EDM) by a spin $\frac{1}{2}$ particle would signal the violation of
time-reversal invariance.  (The existence of a dipole moment signals a
spherically asymmetric distribution of charge.)  Any such effect is likely
small, given the observed magnitude of CP and T violation in the neutral kaon
system, and this makes neutral particles the obvious subject for experiments:
the existence of an electric monopole charge would overwhelm most signals of
the dipole strength.  It is therefore natural to focus on the neutron, which
is the simplest spin-$\frac{1}{2}$ neutral system in nature, and attempts to
measure the neutron's EDM have a long history~\cite{royringo,edmexpt}.

Very accurate measurements are made possible by the ability to produce and
store ultracold ($\lesssim 10\,$eV) polarised neutrons.  Immersing such a
sample in uniform magnetic and electric fields, the existence of a nonzero
neutron EDM, $d_n\neq 0$, manifests itself through a difference between the
precession frequency of the neutron's spin when the magnetic and electric
fields are parallel as compared to when they are antiparallel.  Contemporary
applications~\cite{edmexpt} of this technique yield an upper bound
\begin{equation}
\label{dnub}
|d_n| < 6.3 \times 10^{-26}\,e\,{\rm cm}\; (90\%\,{\rm C.L.}).
\end{equation}
(NB.\ $e/(2 M_n) = 1.0 \times 10^{-14}\,e\,{\rm cm}$.  Therefore, writing
$d_n = e h_n/(2 M)$, where $M$ is the neutron's mass and $h_n$ is its
``gyroelectric ratio,'' Eq.~(\ref{dnub}) corresponds to $|h_n| < 6.0 \times
10^{-12}$.)

Equation~(\ref{dnub}) has proven to be an effective constraint.  For example,
the most general Lagrangian density for a local, renormalisable
colour-$SU(3)$ gauge theory would contain a term:
\begin{equation}
\label{thetaterm}
\frac{\theta}{32\pi^2}\,\varepsilon_{\mu \nu \rho \sigma} \, F_{\mu\nu}^a \,
F_{\rho\sigma}^a\,,
\end{equation}
where $F_{\mu\nu}^a$ is the non-Abelian field strength tensor and $\theta$ is
an undetermined constant - the ``$\theta$ angle.'' However, this term
violates CP invariance and hence generates a nonzero neutron EDM.  Its
strength; i.e., the value of $\theta$, is therefore constrained by
Eq.~(\ref{dnub}), which yields a very low upper
bound~\cite{bruce,meissner,bfhda}
\begin{equation}
|\theta|< (1 \sim 10)\times 10^{-10}\,.
\end{equation}
Currently there is no satisfactory explanation of why this term is so
strongly suppressed and that is the basis of the so-called ``strong CP
problem:'' the goal is to find a reason why $\theta$ should {\it naturally}
be identically zero.  (In the absence of topologically nontrivial gauge field
configurations, Eq.~(\ref{thetaterm}) cannot contribute to the action: it is
a surface term.)

Herein we shall assume $\theta \equiv 0$ so that the phase in the CKM matrix
is the only source of CP and T violation in the standard model.  In this case
a nonzero neutron EDM is proportional to the CKM phase.  (Considerations
relevant for $\theta\neq 0$ are explored, e.g., in Refs.\
\cite{meissner,bfhda}.)

Over many years the experimental upper bound on $d_n$ has steadily decreased
and its small value has also proven very effective in ruling out candidates
for theories that enlarge the standard model.  To illustrate how, we suppose
for the moment that the neutron is a collection of three valence-quarks
described by a symmetric $SU(6)$ spin-flavour wave function.  Then, by
analogy with the magnetic moment,
\begin{equation}
\label{dnCQM}
d_n = \frac{1}{3} ( 4\, d_d - d_u)\,,
\end{equation}
where $d_{u,d}$ are valence-quark EDMs.  In the standard model the first
nonzero contribution to a free quark's EDM appears at third order and
involves a gluon radiative correction (i.e., O$(\alpha_s\,G_F^2)$, for the
same reason that flavour-changing neutral currents are suppressed: the GIM
mechanism) so that~\cite{bruce,dar}
\begin{equation}
\label{ddSM}
d_u < d_d \lesssim 10^{-34}\,e\,{\rm cm}\,.
\end{equation}
Using this in Eq.~(\ref{dnCQM}) gives a result {\it seven}
orders-of-magnitude less than the current experimental bound.  This result is
characteristic: other plausible mechanisms within the standard model, such as
hadronic loop corrections, also give a very small value.  However, the
standard model is peculiar in this regard and candidates for its extension
typically contain many more possibilities for CP and T violation, which {\it
a priori} are not constrained to be small.  Thus Eq.~(\ref{dnub}) is an
important and direct constraint on these extensions because
Eqs.~(\ref{dnCQM}), (\ref{ddSM}) indicate that the standard model
contribution to $d_n$ cannot possibly interfere at a level that could
currently cause confusion.  For example, as our calculation will show, the
viability of the minimal model of spontaneous CP violation~\cite{minSCP},
which involves three Higgs doublets, is endangered by Eq.~(\ref{dnub}).

Our interest in the neutron's EDM stems from a desire to explore the validity
of Eq.~(\ref{dnCQM}) in the sense that, irrespective of the origin of the
valence-quark EDMs, how are they related to the EDM of the bound state?  This
question has recently been explored~\cite{meissner,bfhda,bfhdb} using QCD Sum
Rules~\cite{derek}.  Our analysis, however, will employ instead a recently
developed, well-constrained Poincar\'e covariant, bound-state picture of the
neutron~\cite{regfe,cjbfe,reinhard,jacquesa,jacquesmyriad,cdrqciv} and
therefore affords a necessary complement.

The nonzero neutron charge radius is a clear indication that a symmetric
SU(6) wave function is inadequate for the neutron and there are significant
additional weaknesses.  For example, in making the connection between
Eqs.~(\ref{dnCQM}) and (\ref{ddSM}) no consideration is given to the
necessary momentum-dependence of the dressed-quark mass function, which is a
longstanding prediction of Dyson-Schwinger equation (DSE) studies; e.g.,
Ref.~\cite{cdragw,mr97}, that has recently been confirmed in lattice-QCD
simulations~\cite{latticequark}, nor for confinement and the concomitant
feature that dressed-quarks in the neutron are {\em not} on shell.
(Reference~\protect\cite{cdragw} provides an heuristic guide to DSEs and
References~\protect\cite{revbasti,revreinhard} give an overview of their
contemporary application, with particular emphasis on continuum strong QCD.)

Hitherto, lacking a Poincar\'e covariant bound state picture of the neutron,
these aspects of dressed-quark behaviour and hadron compositeness have only
been explored in studies of the dipole moments of the
$\rho$-meson~\cite{martina,martinb}, a bound state for which sound DSE models
exist.  Assuming an explicit EDM contribution to the photon-current-quark
vertex:
\begin{equation}
\label{EDMvtxa}
d_f\,i\gamma_5 \sigma_{\mu\nu}\, q_\nu \equiv \frac{e}{2
\,m_f}\,e_f\,h_f\,i\gamma_5 \sigma_{\mu\nu}\, q_\nu \,,
\end{equation}
where: $e$ is the positron charge; $m_{f=u,d}$ are the current-quark masses;
$e_{u,d}$, their electric charge fractions; $h_{u,d}$, their gyroelectric
ratios; and $q_\mu$, the momentum transfer, an analogue of Eq.~(\ref{dnCQM})
for a pointlike $\rho$-meson is~\cite{martina}
\begin{equation}
\label{drhoVq}
d_\rho = d_u + d_{\bar d}\,.
\end{equation}
In this case the direct calculations showed that Eq.~(\ref{drhoVq})
underestimates the EDM of a bound state $\rho$-meson, which is composed of a
dressed-quark and a dressed-antiquark, and described by a Bethe-Salpeter
amplitude, by as much as two orders-of-magnitude.  That can materially affect
the inferred constraints on extensions of the standard model.

A quantum field theoretical description of the nuc\-leon as a composite of a
quark and non\-point\-like colour-antitriplet diquark was proposed and
explored in Refs.~\cite{regfe,cjbfe}, and recent studies~\cite{reinhard} have
shown that this approach, based on a Poincar\'e covariant Fadde'ev equation,
is capable of providing a good description of the spectrum of octet and
decuplet baryons.  With this foundation, a product {\it Ansatz} for the
nucleon's Fadde'ev amplitude has been
employed~\cite{jacquesa,jacquesmyriad,cdrqciv} to describe a wide range of
elastic nucleon form factors.  This is the model we use to explore the
implications of Refs.~\cite{martina,martinb} for the neutron's EDM.

In Sec.~\ref{sec:ME} we describe our model in detail, including a discussion
of the form of a CP and T violating coupling of a photon to a dressed-quark.
An analogous dressed-quark-gluon coupling is also admissible~\cite{bruce} and
may yield an equally important contribution to $d_n$~\cite{bfhdb}.  We
neglect it and hence ours is not a complete calculation of $d_n$.
Nevertheless, this and like terms are additive, and their omission does not
qualitatively affect our discussion nor the points we wish to emphasise.  Our
results are presented in Sec.~\ref{sec:R}, and we report more than just the
EDM so as to establish a context for that result and demonstrate the level of
accuracy to be anticipated using our model.  Section~\ref{sec:E} is an
epilogue.

\section{Model Elements}
\label{sec:ME}
\subsection{Dressed-quarks}
\label{subsec:Sp}
The general form of the dressed-quark propagator is~\cite{fn:Euclidean}
\begin{eqnarray}
S(p) & = & -i \gamma\cdot p\, \sigma_V(p^2) + \sigma_s(p^2)\,, \\
     & = & [i\gamma\cdot p\, A(p^2) + B(p^2)]^{-1}\,, \label{SpAB}
\end{eqnarray}
and it is a model-independent result of quark-DSE studies that the wave
function renormalisation and dressed-quark mass:
\begin{equation}
Z(p^2)=1/A(p^2)\,,\;M(p^2)=B(p^2)/A(p^2)\,,
\end{equation}
respectively, exhibit significant momentum dependence for $p^2\lesssim
1\,$GeV$^2$, which is nonperturbative in origin.  (See, e.g.,
Ref.~\cite{pmqciv}.)  This behaviour is a longtime prediction of DSE
studies~\cite{cdragw} and has recently been observed in lattice-QCD
simulations~\cite{latticequark}.  The infrared enhancement of $M(p^2)$ is an
essential consequence of dynamical chiral symmetry breaking (DCSB) and is the
origin of the constituent-quark mass.  With increasing $p^2$ the mass
function evolves to reproduce the asymptotic behaviour familiar from
perturbative analyses, and that behaviour is unambiguously evident for $p^2
\gtrsim 10\,$GeV$^2$~\cite{mr97}.

While numerical solutions of the quark DSE are readily obtained, the utility
of an algebraic form for $S(p)$ when calculations require the evaluation of
numerous multidimensional integrals is self-evident.  With this in mind an
efficacious parametrisation of the dressed-quark propagator, which exhibits
the essential features described above, was introduced in Ref.~\cite{cdrpion}
and has been used extensively in studies of meson properties; e.g.,
Refs.~\cite{cdrpion,mark,kevin,mrpion,mishaSVY,mikea,pichowsky}.  We use it
herein.

The parametrisation is expressed via
\begin{eqnarray}
\label{ssm}
\bar\sigma_S(x) & =&  2\,\bar m \,{\cal F}(2 (x+\bar m^2))\\
&& \nonumber
+ {\cal F}(b_1 x) \,{\cal F}(b_3 x) \,
\left[b_0 + b_2 {\cal F}(\epsilon x)\right]\,,\\
\label{svm}
\bar\sigma_V(x) & = & \frac{1}{x+\bar m^2}\,
\left[ 1 - {\cal F}(2 (x+\bar m^2))\right]\,,
\end{eqnarray}
with $x=p^2/\lambda^2$, $\bar m$ = $m/\lambda$, ${\cal F}(x)=
[1-\exp(-x)]/x$, $\bar\sigma_S(x) = \lambda\,\sigma_S(p^2)$ and
$\bar\sigma_V(x) = \lambda^2\,\sigma_V(p^2)$.  The mass-scale,
$\lambda=0.566\,$GeV, and parameter values
\begin{equation}
\label{tableA} 
\begin{array}{ccccc}
   \bar m& b_0 & b_1 & b_2 & b_3 \\\hline
   0.00897 & 0.131 & 2.90 & 0.603 & 0.185 
\end{array}\;,
\end{equation}
were fixed in a least-squares fit to light-meson observables~\cite{mark}.
The dimensionless $u=d$ current-quark mass in Eq.~(\ref{tableA}) corresponds
to
\begin{equation}
m=5.1\,{\rm MeV}\,.
\end{equation}
($\epsilon=10^{-4}$ in Eq.~(\ref{ssm}) acts only to decouple the large- and
intermediate-$p^2$ domains.)

Dynamical chiral symmetry breaking is expressed in the parametrisation.  It
gives a Euclidean constituent-quark mass
\begin{equation}
M_{u,d}^E = 0.33\,{\rm GeV}, 
\end{equation}
defined~\cite{mr97} as the solution of $p^2=M^2(p^2)$, whose magnitude is
typical of that employed in constituent-quark models~\cite{simon} and for
which the value of the ratio: $M_{u,d}^E/m = 65$, is characteristic and
definitive of light-quarks~\cite{mishaSVY}.  In addition, DCSB is manifest
through a nonzero vacuum quark condensate
\begin{equation}
-\langle \bar qq \rangle_0^{1\,{\rm GeV}^2} = \lambda^3\,\frac{3}{4\pi^2}\,
\frac{b_0}{b_1\,b_3}\,\ln\frac{1\,{\rm GeV}^2}{\Lambda_{\rm QCD}^2} =
(0.221\,{\rm GeV})^3\,,
\end{equation}
where we have used $\Lambda_{\rm QCD}=0.2\,$GeV.  The condensate is
calculated directly from its gauge invariant definition~\cite{mrt97} after
making allowance for the fact that Eqs.~(\ref{ssm}), (\ref{svm}) yield a
chiral-limit quark mass function with anomalous dimension $\gamma_m = 1$.
This omission of the additional $\ln( p^2/\Lambda_{\rm QCD}^2)$-suppression
that is characteristic of QCD is a practical but not necessary
simplification.

Motivated by model DSE studies~\cite{entire}, Eqs.~(\ref{ssm}), (\ref{svm})
express the dressed-quark propagator as an entire function.  Hence $S(p)$
does not have a Lehmann representation, which is a sufficient condition for
confinement~\cite{fn:confinement}.  Employing an entire function for $S(p)$,
whose form is only constrained by the calculation of spacelike observables,
can lead to model artefacts when it is employed directly to calculate
observables involving large timelike momenta~\cite{ahlig}.  An improved
parametrisation is therefore being sought.  Nevertheless, on the subdomain of
the complex plane explored in the present calculation the integral support
provided by an equally efficacious alternative cannot differ significantly
from that of this parametrisation, which explains why we use it herein.

\subsection{Nucleon}
\label{subsec:Nucleon}
The idea that diquark correlations play a significant role in baryon
structure and interactions is almost as old as that of quarks
themselves~\cite{firstdq}.  It was a motivation for the meson-diquark
bosonisation enunciated in Ref.~\cite{regdq}, which provides a picture of
baryons as dressed-quark--diquark composites, and the subsequent
derivation~\cite{regfe} of an homogeneous, Poincar\'e covariant Fadde'ev
equation for baryons that exploits the role of diquark correlations.

Our picture of the nucleon is based on the latter approach.  We represent the
nucleon as a relativistic three-quark bound state, involving a nonpointlike,
Lorentz-scalar diquark correlation, via a product {\it Ansatz} for the
Fadde'ev amplitude:
\begin{eqnarray}
\nonumber \lefteqn{\Psi_3(p_i;\alpha_i,\tau_i)=\epsilon_{c_1 c_2
c_3}\,\Delta^{0^+}(K)\,}\\ 
& & \times \,
[\Gamma_{0^+}(\sfrac{1}{2}p_{[12]};K)]_{\alpha_1\alpha_2}^{\tau_1 \tau_2}\,
[\psi(\ell;P)\,u(P)]^{\tau_3}_{\alpha_3}\,,
\label{Psi3}
\end{eqnarray}
where: $(i\gamma\cdot P + M) u(P)=0$, with $P=p_1+p_2+p_3=:p_{\{123\}}$ the
nucleon's total momentum and $M$ its mass; $\epsilon_{c_1 c_2 c_3}$ is the
Levi-Civita symbol that provides the colour-singlet factor;
$K=p_1+p_2=:p_{\{12\}}$, $p_{[12]}:=p_1-p_2$, $\ell= (p_{\{12\}} - 2 p_3)/3$;
and $(\alpha_i,\tau_i)$ are the quark spinor and isospin labels.

Equation~(\ref{Psi3}) describes the general form of the amplitude in the
scalar diquark subspace.  In this equation $\Delta^{0^+}(K)$ is the
pseudoparticle propagator for a scalar diquark formed from quarks $1$ and
$2$, and $\Gamma_{0^+}$ is a Bethe-Salpeter-like amplitude describing their
relative momentum correlation.  As explained, e.g., in Ref.~\cite{dqandmu},
these functions can be obtained from an analysis of the quark-quark
scattering matrix.  However, following
Refs.~\cite{jacquesa,jacquesmyriad,cdrqciv}, for simplicity herein we employ
parametrisations:
\begin{eqnarray}
\label{dqprop}
\Delta^{0^+}(K^2) & = & \frac{1}{m_{0^+}^2}\,{\cal
F}(K^2/\omega_{0^+}^2)\,,\\
\label{gdq}
\Gamma_{0^+}(k;K) & = & \frac{1}{{\cal N}_{0^+}} C\, i\gamma_5\, i\tau_2\,
{\cal F}(k^2/\omega_{0^+}^2)\,,
\end{eqnarray}
where $C=\gamma_2\gamma_4$ is the charge conjugation matrix and ${\cal
N}_{0^+}$ is a calculated, canonical normalisation constant that ensures,
e.g., that a $(ud)$-diquark has electric charge fraction $(1/3)$ for
$K^2=-m_{0^+}^2$.  The parameters are a width, $\omega_{0^+}$, and a
pseudoparticle mass, $m_{0^+}$, which have ready physical interpretations:
the length $r_{0^+}=1/\omega_{0^+}$ is a measure of the mean separation
between the quarks in the scalar diquark; and the distance $l_{qq}= 1/m_{qq}$
represents the range over which a true diquark correlation in this channel
can persist {\it inside} a baryon.  (NB.\ The absence of a particle-like
singularity in the pseudoparticle propagator presented in Eq.~(\ref{dqprop})
is sufficient to ensure that the diquark is confined inside the
baryon~\cite{fn:confinement}.)

The remaining element in Eq.~(\ref{Psi3}), $\psi$, is a Bethe-Salpeter-like
amplitude that describes the relative momentum correlation between the
dormant quark and the diquark's centre-of-momentum.  Using
Eqs.~(\ref{dqprop}), (\ref{gdq}), $\psi$ can be obtained by solving a
Poincar\'e covariant Fadde'ev equation for the nucleon~\cite{regfe}:
\begin{eqnarray}
\nonumber \psi(k;P)\,u(P) & = & -2 \int\frac{d^4
\ell}{(2\pi)^4}\,\Delta^{0^+}(K_\ell)\,\Gamma_{0^+}(k+\ell/2;K) \\ 
\nonumber & & \times \, S(\ell_{\rm ex})^{\rm
T}\,\bar\Gamma_{0^+}(\ell+k/2;-K_k)\,\\
& & \times\, S(\ell_1) \,\psi(\ell;P)\,u(P) \,,
\label{faddeev}
\end{eqnarray}
where $K_\ell= -\ell + (2/3)P$, $\ell_{\rm ex}= -\ell-k-P/3$, $\ell_1=\ell +
P/3$.  For a positive energy nucleon, the solution has the general form
\begin{equation}
\label{psi3}
\psi(\ell;P)= f_1(\ell;P) 1\!\rule{0.3ex}{1.55ex} - \frac{1}{M} \left( i
\gamma\cdot \ell - \ell\cdot \hat P \, 1\!\rule{0.3ex}{1.55ex}\right)
f_2(\ell;P)\,, 
\end{equation}
where $\hat P^2 = -1$ and, in the nucleon rest frame, $f_{1,2}$ describe,
respectively, the upper$/$lower component of the dressed-nucleon spinor.

To learn about $\psi$ we solved Eq.~(\ref{faddeev}) and also its extension to
include an axial-vector diquark correlation, a sound foundation for the
latter step having been laid in Ref.~\cite{reinhard}.  We used the
dressed-quark propagator described in Sec.~\ref{subsec:Sp} and, to further
simplify and so expedite the calculations, we retained only the first
Chebyshev moment of the functions $f_{1,2}$ in Eq.~(\ref{psi3}); i.e., we
assumed $f_{1,2}(\ell;P)=f_{1,2}(\ell^2;P^2)$.  In this way we
found~\cite{cdrqciv} that a simultaneous description of the nucleon and
$\Delta$ masses is possible when the axial-vector diquark correlations are
included.  That description requires (in GeV)
\begin{equation}
\label{faddeevfit}
\begin{array}{cccc|cc}
\omega_{0^+} & m_{0^+} & \omega_{1^+} & m_{1^+} & M_N & M_\Delta \\\hline
0.42 & 0.64 & 1.09 & 0.86 & 0.94 & 1.23 
\end{array}\,,
\end{equation}
where $\omega_{1^+}$, $m_{1^+}$ are obvious analogues in the axial-vector
channel of the scalar diquark parameters in Eqs.~(\ref{dqprop}), (\ref{gdq}),
and corresponds to (in fm) $r_{0^+}= 0.47$, $r_{1^+}= 0.23$, $l_{0^+}=0.31$,
$l_{1^+}=0.18$.  (The last two columns in Eq.~(\ref{faddeevfit}) are the
calculated masses.  A description of the $\Delta$, for which $J=3/2$, is
obviously impossible with only a scalar diquark.)  In our calculation the
value of $m_{0^+}$ was taken from Refs.~\cite{jacquesa,jacquesmyriad} and
that of the ratio $m_{0^+}/m_{1^+}=0.78$ from the Bethe-Salpeter equation
studies of Ref.~\cite{conradsep}, which is consistent with that obtained in
lattice-QCD simulations~\cite{latticediquark}.  The parameters
$\omega_{0^+,1^+}$ were then varied to fit $M_{N,\Delta}$.  The observations:
$m_{0^+}/m_{1^+}=0.78 \approx 0.76 = M_N/M_\Delta$; $M^E_{u,d} + m_{0^+} =
0.97$; and $M^E_{u,d} + m_{1^+} = 1.19$, are suggestive.  However, our own
and other~\cite{reinhard,reinharddelta} Fadde'ev equation studies show that
the nucleon contains a significant axial-vector diquark component (neglecting
it, the nucleon mass is $\sim 40$\% too large~\cite{cdrqciv}) and hence the
origin of the $N$ and $\Delta$ masses and mass splitting does not lie simply
in summing over constituent masses.

For the purpose of developing an intuitive understanding, our eigenvector
solution for the nucleon can adequately be approximated as~\cite{cdrqciv}
\begin{equation}
\label{psi3model}
\psi(\ell;P)= \frac{1}{{\cal N}_{\psi}}\,{\cal F}(\ell^2/\omega_\psi^2) \left[
1\!\rule{0.3ex}{1.55ex} - \frac{\mbox{\sc r}}{M} \left( i
\gamma\cdot \ell - \ell\cdot \hat P \, 1\!\rule{0.3ex}{1.55ex}\right)\right]
\end{equation}
with calculated values of $\omega_\psi \approx 0.4\,$GeV ($r_\psi =
0.49\,$fm) and $\mbox{\sc r}\approx 0.5$.  This makes clear that
$r_\psi\gtrsim r_{0^+}\gtrsim r_{1^+}$, which is a necessary condition for
the internal consistency of the quark$+$diquark Fadde'ev equation description
since it signifies that the mean separation between the quarks in the
constituent diquarks is no more than the size of the nucleon.  Furthermore,
the value of $\mbox{\sc r}$ indicates that the lower component is a
significant piece of a relativistic nucleon's spinor.  (${\cal N}_\psi$ is a
calculated normalisation constant that ensures, e.g., that the proton has
unit charge.)

Following this study we can complete the specification of a well-informed
product {\it Ansatz}.  Equation~(\ref{Psi3}), with Eqs.~(\ref{dqprop}),
(\ref{gdq}), (\ref{psi3model}), provides a two-parameter model: we fix the
values of $\mbox{\sc r}=0.5$ and $m_{0^+}\approx 0.64\,$GeV, motivated by the
Fadde'ev equation studies, and allow $\omega_{0^+}$ and $\omega_\psi$ to vary
so as to obtain a least-squares fit to the proton's electric form factor, as
described in Refs.~\cite{jacquesa,jacquesmyriad,cdrqciv}.  With this
expedient we have an algebraic scalar diquark model that provides accurate
estimates of known observables, as we show in Sec.~\ref{sec:R}, and hence can
easily be used to obtain realistic constraints on the neutron's EDM that
reflect the influence of those aspects of strong QCD that we identified in
Sec.~\ref{introduction}: DCSB; quark confinement; and hadron compositeness.

\subsection{Quark-Photon Coupling}
\label{subsec:Gamma}
A calculation of the electromagnetic interaction of a composite particle
cannot proceed without an understanding of the coupling between the photon
and the bound state's constituents.  This is illustrated with particular
emphasis in Refs.~\cite{cdrpion,mrpion,photonanomaly}, which consider effects
associated with the Abelian anomaly.  When quarks are dressed as described in
Sec.~\ref{subsec:Sp}, only a dressed-quark-photon vertex, $\Gamma_\mu$, can
satisfy the vector Ward-Takahashi identity:
\begin{equation}
\label{vwti}
q_\mu \, i\Gamma_\mu(\ell_1,\ell_2) = S^{-1}(\ell_1) -
S^{-1}(\ell_2)\,,
\end{equation}
where $q=\ell_1-\ell_2$ is the photon momentum flowing into the vertex.  The
constraints that this identity and other features of a renormalisable quantum
field theory place on the form of $\Gamma_\mu$ have been explored extensively
in Refs.~\cite{ayse}.

$\Gamma_\mu$ is the solution of an inhomogeneous BSE, and the pointwise
behaviour of the solution has been elucidated in the numerical studies of
Refs.~\cite{pieterGamma}.  However, for our purposes we again prefer an
efficacious algebraic parametrisation and that is provided by~\cite{bc80}
\begin{eqnarray}
\nonumber \lefteqn{i\Gamma_\mu(\ell_1,\ell_2) = 
i\Sigma_A(\ell_1^2,\ell_2^2)\,\gamma_\mu +
(\ell_1+\ell_2)_\mu\,}\\ 
& & \label{bcvtx}
\times \left[\sfrac{1}{2}i\gamma\cdot (\ell_1+\ell_2) \,
\Delta_A(\ell_1^2,\ell_2^2) + \Delta_B(\ell_1^2,\ell_2^2)\right]\,;\\
&&  \Sigma_F(\ell_1^2,\ell_2^2) = \sfrac{1}{2}\,[F(\ell_1^2)+F(\ell_2^2)]\,,\\
&& \Delta_F(\ell_1^2,\ell_2^2) =
\frac{F(\ell_1^2)-F(\ell_2^2)}{\ell_1^2-\ell_2^2}\,,
\label{DeltaF}
\end{eqnarray}
where $F= A, B$; i.e., the scalar functions in Eq.~(\ref{SpAB}).  A feature
of Eq.~(\ref{bcvtx}) is that $\Gamma_\mu$ is completely determined by the
dressed-quark propagator.  Furthermore, when using the dressed-quark
propagator parametrisation, improvements on this vertex {\it Ansatz} can only
modify our results by $\lesssim 10\,$\%, as illustrated, e.g., in
Refs.~\cite{pichowsky,piloop}.

Equations~(\ref{bcvtx})-(\ref{DeltaF}) and Refs.~\cite{ayse,pieterGamma,bc80}
consider only the CP preserving part of the dressed-vertex.  When the
possibility of CP and T violation is admitted, additional contributions are
possible and the form that has most often been considered is that of
Eq.~(\ref{EDMvtxa}):
\begin{equation}
\label{EDMvtxaa}
\Gamma_\mu^-(\ell_1,\ell_2) = 
\frac{e}{2 m}\,Q\,H^-(q^2)\,\gamma_5 \sigma_{\mu\nu}\, q_\nu\,,
\end{equation}
where: $e$ is the positron charge, $m$ is the $u=d$ current-quark mass;
$Q={\rm diag}[2/3,-1/3]$ is the quark charge fraction isospin matrix; and
$H^-={\rm diag}[h_u^-(q^2),h_d^-(q^2)]$ defines an analogous quark EDM
matrix, wherein we acknowledge that the quarks' EDMs may be
momentum-dependent.  Plainly, adding this term preserves Eq.~(\ref{vwti}).

As discussed in Ref.~\cite{martinb}, another contribution is possible:
\begin{eqnarray}
\nonumber \Gamma_\mu^+(\ell_1,\ell_2) & = & \frac{e}{2
\,m}\,Q\,H^+(q^2)\,i\gamma_5\\
& & \times \, \left[(\ell_1+\ell_2)_\mu - (\ell_1 + \ell_2) \cdot \hat{q}
 \,\hat{q}_\mu \right] ,
\end{eqnarray}
where: $\hat q^2 = 1$; $H^+(q^2)$ is an obvious analogue of $H^-(q^2)$; and
adding this term also preserves Eq.~(\ref{vwti}).  Note though that in any
calculation where the quarks are assumed to be on-shell, and hence described
by spinors for which $(i\gamma\cdot \ell + m) u(\ell)=0$, then
\begin{equation}
\label{gordonI}
i \bar u(\ell_1) \,\gamma_5\,(\ell_1+\ell_2)_\mu \, u(\ell_2)
= \bar u(\ell_1)\, \gamma_5 \, \sigma_{\mu\nu}\, q_\nu \,u(\ell_2)\,,
\end{equation}
and consequently the two structures are equivalent.  However,
Eq.~(\ref{gordonI}) is not satisfied by the dressed-quarks in the bound state
nucleon: they are confined and may not even have a mass shell, so that in the
general case the two contributions are distinguishable and must be treated
separately.  As an example, for the $\rho$-meson the different operator
structures generate individual contributions that differ by $\lesssim 20$\%
from their average value~\cite{martinb}.

In our calculations then we employ the following algebraic {\it Ansatz} for
the dressed-quark-photon vertex
\begin{equation}
\label{vtxTotal}
 \Gamma_\mu^Q(\ell_1,\ell_2)= 
 Q\,\Gamma_\mu(\ell_1,\ell_2)
+ \Gamma_\mu^+(\ell_1,\ell_2) 
+ \Gamma_\mu^-(\ell_1,\ell_2) \,.
\end{equation}

\section{Results}
\label{sec:R}
The electromagnetic nucleon current is
\begin{eqnarray}
\label{Jnucleon}
J_\mu(P^\prime,P) & = & ie\,\bar u(P^\prime)\, \Lambda_\mu(q,P) \,u(P)\,, \\
& = & \nonumber i e \,\bar u(P^\prime)\,\left( \gamma_\mu F_1(q^2) +
\frac{1}{2M}\, \sigma_{\mu\nu}\,q_\nu\,F_2(q^2) \right. \\
& & + \left. \frac{1}{2 M}\, \gamma_5\sigma_{\mu\nu}\,q_\nu\,{\cal H}(q^2)
\right) u(P)\,,
\label{JnucleonB}
\end{eqnarray}
where: the spinors satisfy $\gamma\cdot P \, u(P) = i M u(P)$, $\bar u(P)\,
\gamma\cdot P = i M \bar u(P)$, with $M=0.94\,$GeV; $R=P^\prime + P$ and
$q\cdot P =0$; and $\Lambda_\mu$ is the nucleon-photon vertex.  $F_1$ and
$F_2$ are the usual Dirac and Pauli electromagnetic form factors of the
nucleon, in terms of which the electric and magnetic form factors are
\begin{eqnarray}
G_E(q^2) & = & F_1(q^2) - \frac{q^2}{4 M^2} F_2(q^2)\,,\\
G_M(q^2) & = & F_1(q^2) + F_2(q^2)\,.
\end{eqnarray}
The remaining term yields ${\cal H}$ and hence the nucleon's EDM form factor,
which vanishes in the absence of a CP and T violating quark-photon coupling.
Equation~(\ref{JnucleonB}) is the general form because Eq.~(\ref{gordonI}) is
valid for an on-shell nucleon.

With the specification of the elements in Sec.~\ref{sec:ME}, an impulse
approximation calculation of the electromagnetic properties of the nucleon is
straightforward using the expression for $\Lambda_\mu$ given in the appendix.
Each calculation requires the evaluation of a number of multidimensional
integrals, which we accomplish using Monte-Carlo methods, requiring a
statistical accuracy of $\lesssim 1\,$\%.  

\begin{table}
\caption{\label{tableB} Calculated values of a range of well-known physical
observables.  The ``Obs.''\ column reports experimental values~\cite{expt} or
values employed in a typical meson exchange model~\protect\cite{harry}.  The
remaining columns report our results, obtained using the Fadde'ev {\it
Ansatz} parameters in Eq.~(\protect\ref{Sets}).}
\vspace*{1ex}
\begin{ruledtabular}
\begin{tabular*}
{\hsize}
{l@{\extracolsep{0ptplus1fil}}c@{\extracolsep{0ptplus1fil}}c@{\extracolsep{0ptplus1fil}}c}
%{l|c|c|c}
                  & Obs.         & Set~1   & Set~2 \\ \hline
$(r_p)^2\,($fm$^2$) & $(0.87)^2$ & $(0.78)^2$ & $(0.83)^2$ \\
$(r_n)^2\,($fm$^2$) & $-(0.34)^2$& $-(0.40)^2$ & $-(0.33)^2$ \\
$\mu_p \, (\mu_N)$  & $2.79$     & $2.83$   & $2.82$ \\
$\mu_n \, (\mu_N)$  &$-1.91$     &$-1.61$   &$-1.62$ \\
$\mu_n/\mu_p$       &$-0.68$     &$-0.57$   &$-0.57$ \\
$g_{\pi NN}$        &$13.4$      &$13.9$   &$15.4$ \\
$\langle r_{\pi NN}^2\rangle\,($fm$^2$)
                    & $(0.93$-$1.06)^2$  &$(0.63)^2$   &$(0.69)^2$ \\
$g_A$               &$1.26$    &$0.98$   &$1.27$ \\
$\langle r_{A}^2\rangle\,($fm$^2$)
                    &$(0.68\pm 0.12)^2$    &$(0.83)^2$   &$(0.77)^2$ \\
$g_{\rho NN}$       &$6.4$    &$6.66$   & $8.60$ \\
$f_{\rho NN}$       &$13.0$    &$15.8$   &$16.7$ \\
$g_{\omega NN}$     &$7$--$10.5$ & $12.2$   & $15.5$ \\
$f_{\omega NN}$     &           & $6.5$   & $5.2$ 
\end{tabular*}
\end{ruledtabular}
\end{table}

Our results for an illustrative range of well-known quantities are presented
in Table~\ref{tableB}, which serves to exemplify the accuracy of our scalar
diquark model of the nucleon.  To display the sturdiness of the model we used
two parameter sets (dimensioned quantities in GeV)
\begin{equation}
\label{Sets}
\begin{array}{l|cc|cc}
            & \mbox{\sc r} & m_{0^+} & \omega_{0^+} & \omega_\psi\\\hline
{\rm Set~1} & 0.5 & 0.62 & 0.79 & 0.23 \\
{\rm Set~2} & 1.0 & 0.63 & 0.59 & 0.18
\end{array}
\end{equation}
with, in each case, the values of $\omega_{0^+}$ and $\omega_{\psi}$
determined via a least-squares fit to the proton's electric form factor.
Set~1 is ``preferred'' because the value of {\sc r} is fixed to that obtained
in the Fadde'ev equation study described in Sec.~\ref{subsec:Nucleon}.  (The
minor adjustment of $m_{0^+}$ is necessary to preserve neutrality of the
neutron subsequent to the variation of $\omega_{0^+}$, $\omega_{\psi}$, see
next paragraph.)  In comparison with the form factors depicted in
Refs.~\cite{jacquesa,jacquesmyriad}, both of which used $\mbox{\sc r}=0$ and
hence overlooked an important qualitative outcome of Fadde'ev equation
studies, our calculated $G^n_E(q^2)$ is much improved cf.\ the data, as seen,
e.g., in Ref.~\cite{cdrqciv}.  The other form factors are little affected.
(NB.\ Only the electromagnetic properties reported in Table~\ref{tableB} can
be calculated directly from the information provided herein.  The remaining
quantities were calculated as described in Ref.~\protect\cite{jacquesmyriad}:
$\pi NN$ -- using Eq~(22) of that reference; vector meson -- using Eq.~(38);
and axial form factor -- using Eq.~(60).)

We remark that there is a quantitative discrepancy between the values of
$\omega_{0^+}$, $\omega_\psi$ obtained here and those obtained in the
Fadde'ev equation study, Eqs.~(\ref{faddeevfit}), (\ref{psi3model}).  That is
to be anticipated because here we are requiring a fit to wide range
observables in a circumscribed model space; i.e., the discrepancy is an
artefact of our scalar-diquark product {\it Ansatz}.  However, our model's
simplicity and illustrative utility outweigh the defect because in this and
its earlier applications the qualitative implications of the defect are
readily identifiable.

\subsection{Electric dipole moment}
Having established a context, we now report our results for the EDM.  Using
Eq.~(\ref{vtxTotal}) one finds
\begin{equation}
\label{calH}
{\cal H} = \sfrac{M}{m}\, {\rm diag}\left(
\sum_{\stackrel{\sigma=+,-}{f=u,d}}\,w_f^\sigma(p)\,e_f\,h_f^\sigma,
\sum_{\stackrel{\sigma=+,-}{f=u,d}}\,w_f^\sigma(n)\,e_f\,h_f^\sigma
\right),
\end{equation}
an isospin matrix, where the calculated values of $w_f^\sigma(N)$ express the
dependence of the nucleon's EDM on its internal structure.  Clearly, because
of isospin symmetry,
\begin{equation}
w_u^\sigma(p) = w_d^\sigma(n)\,,\;w_d^\sigma(p) = w_u^\sigma(n)\,.
\end{equation}
The $M/m$ multiplicative factor in Eq.~(\ref{calH}) makes explicit the
importance of the DCSB mechanism in amplifying the contribution from
dressed-quarks to the nucleon's EDM.  

Existing experiments constrain ${\cal H}(q^2=0)$ and in Table~\ref{tableC} we
report the relevant values of $w_f^\sigma(n)$.  (There are no cancellations
between contributions from different $\Lambda_\mu^i$,
Eqs.~(\protect\ref{nucvtx})-(\protect\ref{L5}).  The values are comparable in
magnitude to their analogues obtained using QCD Sum Rules, e.g.,
Ref.~\cite{bfhdb}.)  The table makes clear the extent to which
Eq.~(\ref{gordonI}) (the on-shell assumption) is violated by the
dressed-quarks in the nucleon bound state: for the doubly represented quark
flavour the difference is quantitatively similar to that in the
$\rho$-meson~\cite{martinb}.  The difference is much larger for the odd
flavour quark.  However, that is likely an artefact of only retaining a
scalar diquark, in which case $\Lambda_\mu^3$ in Eq.~(\ref{Lambda3}) provides
a sole, unmatched contribution to $h^\pm$.  On the information available we
therefore judge that the magnitude of this effect is best estimated from the
difference for the doubly represented flavour.

\begin{table}
\caption{\label{tableC} Calculated values of the coefficients in
Eq.~(\protect\ref{calH}).  (All quantities are dimensionless.)  In the
absence of dressed-quark confinement and off-shell effects, the entries in
the two rightmost columns would be zero.  The parameter values for Sets~1 and
2 are given in Eq.~(\protect\ref{Sets}).}
\vspace*{1ex}
\begin{ruledtabular}
\begin{tabular*}
{\hsize}
{l@{\extracolsep{0ptplus1fil}}c@{\extracolsep{0ptplus1fil}}c@{\extracolsep{0ptplus1fil}}c@{\extracolsep{0ptplus1fil}}c@{\extracolsep{0ptplus1fil}}c@{\extracolsep{0ptplus1fil}}c}
        & $w^+_d(n)$ & $w^+_u(n)$ & $w^-_d(n)$ & $w^-_u(n)$ 
        & $\frac{w_d^--w_d^+}{w_d^-+w_d^+}$ 
        & $\frac{w_u^--w_u^+}{w_u^-+w_u^+}$ \\ \hline
Set~1   & $0.560$ & $0.055$ & $0.756$ & $-0.111$ & 0.15 & 2.9 \\
Set~2   & $0.647$ & $0.043$ & $0.878$ & $-0.102$& 0.15 & 2.5 
\end{tabular*}
\end{ruledtabular}
\end{table}

Using Eqs.~(\ref{tableA}), (\ref{calH}) and the Set~1 results from
Table~\ref{tableC} we obtain
\begin{eqnarray}
\nonumber
h_n  & = & 185.2 \, \sfrac{1}{3} \left( - 0.560 \,h_d^+ + 0.110\,h_u^+ \right.\\
& & \left. - 0.756\,h_d^- - 0.222\,h_u^- \right)\,, \label{hn} \\
\nonumber h_p & = & 185.2 \, \sfrac{1}{3} \left( -0.055 \, h_d^+ +
0.111\,h_d^- \right. \\
& & \left. + 1.12 \, h_u^+ + 1.51\,h_u^-\right)\,.
\label{hp} 
\end{eqnarray}
The ratio of the coefficients of $h_d^-$ and $h_u^-$ in Eq.~(\ref{hn}) is
$3.4$, while the value of this ratio obtained from Eq.~(\ref{dnCQM}) is
$2.0$.  The difference between these values gauges the extent to which
$SU(6)$ spin-flavour symmetry is broken by quark$\,+\,$quark$\,=\,$diquark
clustering in the nucleon bound state.  This difference may change a little
but will not vanish upon the inclusion of axial-vector diquark correlations
in the nucleon's Fadde'ev amplitude.  (The limited effect that axial-vector
diquarks have on other observables that do not involve cancellations between
the $\Lambda_\mu^i$-contributions~\cite{cdrqciv,reinharddelta} makes us
confident of this.)  The $h^+$ contributions do not have an analogue
perturbatively nor in the constituent quark model.

We observe that hadronic loop insertions are unambiguously an additive
correction to our impulse approximation and, based on analyses of other
observables~\cite{cdrqciv,kevin,pichowsky,piloop,rhopipi}, we expect that for
realistic hadron masses their contribution will modify Eqs~(\ref{hn}),
(\ref{hp}) by $\lesssim 10\,$\%.  This suppression is due to the
compositeness of the hadrons: away from the position of well-known kinematic
singularities, the need to extrapolate bound state properties
off-shell~\cite{fn:offshell}, for the evaluation of integrals over spacelike
momenta; and the integration cutoffs applied by the hadrons' finite size,
both lead to a quenching of the loop contributions.

If we assume that CP-violating current-quark-level interactions yield $h_d\gg
h_u$, as is the case, e.g., in Higgs boson exchange models~\cite{martinb},
then
\begin{equation}
\label{factor}
h_n = - 40.6 \,\left(0.85\,h_d^+ + 1.15\,h_d^-\right)
\end{equation}
and Eq.~(\ref{dnub}) applies the following bound on the $d$-current-quark
gyroelectric ratios:
\begin{equation}
\label{hdU}
|0.85\,h_d^+ + 1.15\,h_d^-| < 14.7 \times 10^{-14}\,.
\end{equation}
Using Eq.~(\ref{hp}) it is clear that, for $h_d\gg h_u$, $d_p \lesssim 0.1\,
d_n$.

The DCSB mechanism, which is responsible for turning the current-quark mass
into the constituent-quark mass, as reviewed in Sec.~\ref{subsec:Sp}, is
responsible for the suppression of $h_d$ with respect to $h_n$ by the factor
of $\sim 40$ in Eq.~(\ref{factor}).  This value is a lower bound on the
magnitude of the effect.  It will increase when using any reasonable {\it
Ansatz} that incorporates strong interaction dressing in $\Gamma_\mu^{\pm}$,
analogous to that described by $\Sigma_F$, $\Delta_F$ in the CP and T
preserving part of the vertex.  We can estimate the scale of this effect by
comparing $\omega_u^-(p)(=\omega_d^-(n))$ and $\omega_d^-(p)(=\omega_u^-(n))$
with lattice estimates~\cite{tensor} of the proton's tensor charges: $\delta
u=0.839(60)$, $\delta d= -0.180(10)$.  It is thus apparent that
Eq.~(\ref{hdU}) overestimates the upper bound by less than a factor of two.

The result in Eq.~(\ref{ddSM}) corresponds to $|h_d^++h_d^-|\lesssim
10^{-22}$, so Eq.~(\ref{hdU}) does not challenge the standard model.
However, with~\cite{epe} Re$(\epsilon^\prime/\epsilon)= (2.1 \pm 0.46)\times
10^{-3}$, the calculation in Sec.~VI of Ref.~\cite{bruce99} corresponds to a
Weinberg model~\cite{minSCP} prediction of~\cite{bruce01}
\begin{equation}
|h_d^+ + h_d^- | = (0.3 \sim 9.0 )\times 10^{-14}\,.
\end{equation}
Hence this model is threatened by our estimate of the bound, Eq.~(\ref{hdU}),
which incorporates a well-constrained modelling of the effects of DCSB, and
the binding and confinement of dressed-quarks in the neutron.  

We emphasise that the results in Table~\ref{tableC} are independent of the
means used to calculate $h_f^\sigma$.  Hence Eq.~(\ref{hn}) can be applied to
constrain any extension of the standard model.  Here we have only exemplified
that using a particular candidate.

\section{Epilogue}
\label{sec:E}
The results are clear.  Quantitatively -- Our value for $d_n$ will be
modified by the well-constrained inclusion of axial-vector diquark
correlations.  However, based on the effect this has on other observables,
the correction should not exceed $15\,$\%.  Qualitatively -- $1)$ The scale
of dynamical chiral symmetry breaking in QCD generates a significant
amplification of the contribution from a current-quark's EDM to that of the
bound state containing it: the minimal enhancement factor is roughly the
ratio of the constituent-quark and current-quark masses, and is
understandable and calculable through the necessary momentum-dependence of
the dressed-quark mass function.  $2)$ Confinement and compositeness entail
that the dressed-quarks comprising the bound state are not on shell and hence
the two CP and T violating operator structures that are indistinguishable in
the free-quark limit yield materially different contributions to the EDM of
the bound state.  As a consequence these operator structures must be analysed
and their strengths determined independently in any model that provides for
CP and T violation.  Both these effects should be accounted for in using
$d_n$ as a means of constraining extensions of the standard CKM model.  These
conclusions bear equally on the effects of operator structures we have
neglected.

\appendix
\section{Impulse approximation}
As made clear in Sec.~2.3 of Ref.~\cite{revbasti}, when using an
antisymmetrised product {\it Ansatz} for the nucleon's Fadde'ev amplitude the
impulse approximation is 
\begin{eqnarray}
\label{nucvtx}
\Lambda_\mu(q,P) &=&
\Lambda^1_\mu(q,P)
+ 2 \sum_{i=2}^5\,\Lambda^i_\mu(q,P)\,,
\end{eqnarray}
where: 
\begin{eqnarray}
\nonumber \Lambda^1_\mu(q,P) & = & 3 \int\!\!\sfrac{d^4 \ell}{(2\pi)^4}\,
\psi(\ell- \sfrac{2}{3}q;P) \,\Delta^{0^+}(K)\,\\
& & \times\,\psi(\ell;P)\,\Lambda_\mu^S(p_3+q,p_3) \,,
\label{L1}
\end{eqnarray}
with
$K=\ell + \sfrac{2}{3} P$, 
$p_3= \sfrac{1}{3} P - \ell$, 
$\Lambda_\mu^S(\ell_1,\ell_2) =
S(\ell_1)\,\Gamma_\mu^Q(\ell_1,\ell_2)\,S(\ell_2)$; and 
\begin{eqnarray}
\nonumber \Lambda^2_\mu(q,P) & = & 6\int\!\!\sfrac{d^4 k}{(2\pi)^4}\sfrac{d^4
\ell}{(2\pi)^4}\, \Omega(p_1+q,p_2,p_3)\,\Omega(p_1,p_2,p_3) \\
&& \times \, {\rm tr}_{DF}\left[\Lambda_\mu^S(p_1+q,p_1) S(p_2)\right]\,
S(p_3)\,,
\end{eqnarray}
where
$p_1= \sfrac{1}{2} K + k$, 
$p_2= \sfrac{1}{2} K - k$, 
$6 = \varepsilon_{c_1 c_2 c_3} \varepsilon_{c_1 c_2 c_3}$ is the colour
contraction, and
\begin{eqnarray}
\nonumber \Omega(p_1,p_2,p_3) & = & \Delta^{0^+}(p_{\{12\}}) \,
\Gamma_{0^+}(\sfrac{1}{2}p_{[12]};p_{\{12\}})\, \\
& & \times \, 
\psi(\sfrac{1}{3}[p_{\{12\}}-2p_3];P) \,.
\end{eqnarray}
$\Lambda_\mu^2$ describes the photon probing the structure of the scalar
diquark correlation, and contributes equally to both the proton and neutron.
That contribution is trivially zero for the EDM.  This merely reflects the
fact that such a moment is forbidden to a scalar particle, so the only
nontrivial contribution is that of the diquark's CP- and T-preserving
electromagnetic form factor.  The remaining terms are
\begin{eqnarray}
\nonumber \Lambda^3_\mu(q,P) & = & 6\int\!\!\sfrac{d^4 k}{(2\pi)^4}\sfrac{d^4
\ell}{(2\pi)^4}\, \Omega(p_1,p_2,p_3)\, \\
&& \nonumber \times\, \Omega(p_1+q,p_3,p_2)\,S(p_2) \, (i\tau_2)^{\rm T}\\
&& \times\, \Lambda^S_\mu(p_1,p_1+q)\,(i \tau_2)\,S(p_3)\,,\\
\label{Lambda3}
\nonumber \Lambda^4_\mu(q,P) & = & 6\int\sfrac{d^4 k}{(2\pi)^4}
\sfrac{d^4 \ell}{(2\pi)^4}\,\Omega(p_1,p_3,p_2+q)\, \\ 
&& \nonumber \times \,\Omega(p_1,p_2,p_3) \,\Lambda^S_\mu(p_2+q,p_2) \\
&& \times \,S(p_1) \, S(p_3)\,,\\
\nonumber \Lambda^5_\mu(q,P) & = & 6\int\sfrac{d^4
k}{(2\pi)^4}\sfrac{d^4 \ell}{(2\pi)^4}\, \Omega(p_1,p_3+q,p_2)\, \\ 
&& \nonumber \times \,\Omega(p_1,p_2,p_3)\, S(p_2)\,S(p_1)\,\\
&& \times \, \Lambda_\mu^S(p_3+q,p_3)
\,.
\label{L5}
\end{eqnarray}

Of these five terms, the only $u$-quark contribution to the neutron EDM comes
from $\Lambda^3_\mu$ -- the other four terms sum the contribution from the
$d$-quark.  (The situation is reversed for the proton.) 

Our numerical results are obtained by evaluating these integrals using
Monte-Carlo methods and the input specified in Eqs.~(\ref{ssm}), (\ref{svm}),
(\ref{dqprop}), (\ref{gdq}), (\ref{psi3model}) and (\ref{vtxTotal}).

\begin{acknowledgments}
We acknowledge interactions with J.C.R.\ Bloch.  This work was supported by
the US Dept. of Energy, Nuclear Physics Division, under contract
no.~\mbox{W-31-109-ENG-38}.  SMS acknowledges financial support from the
Deutsche Forschungsgemeinschaft (SCHM 1342/3-1).
\end{acknowledgments}

%\bibliography{edm}

\begin{thebibliography}{99}
\bibitem{fn:Euclidean} In our Euclidean formulation: $p\cdot q=\sum_{i=1}^4
p_i q_i$; $\{\gamma_\mu,\gamma_\nu\}=2\,\delta_{\mu\nu}$; $\gamma_\mu^\dagger
= \gamma_\mu$; $\sigma_{\mu\nu}= \sfrac{i}{2}[\gamma_\mu,\gamma_\nu]$; and
tr$_D[\gamma_5\gamma_\mu\gamma_\nu\gamma_\rho\gamma_\sigma]=
-4\,\epsilon_{\mu\nu\rho\sigma}$, $\epsilon_{1234}= 1$.
\bibitem{fn:confinement} This is a sufficient condition for confinement
because of the associated violation of reflection positivity, as discussed,
e.g., in Sec.~6.2 of Ref.~\cite{cdragw}, Sec.~2.2 of Ref.~\cite{revbasti} and
Sec.~2.4 of Ref.~\cite{revreinhard}.
\bibitem{fn:offshell} The ambiguity inherent in this extrapolation is
exemplified in Sec.~2.3 of Ref.~\cite{revbasti} and the only unambiguous
procedure is described in Sec.~VI of Ref.~\cite{jacquesmyriad}.
\end{thebibliography}

\end{document}